\documentclass[twocolumn,showpacs,preprintnumbers,amsmath,amssymb]{revtex4}
\usepackage{graphicx}% Include figure files
\usepackage{dcolumn}% Align table columns on decimal point
\usepackage{bm}% bold math
\begin{document}
%\draft

\title{Electronic Phase Separation in  Manganite/Insulator Interfaces.}

\author{Luis Brey }

\affiliation{\centerline {Instituto de Ciencia de Materiales de
Madrid (CSIC),~Cantoblanco,~28049~Madrid,~Spain.}} \
%\maketitle
\begin{abstract}

By using a realist microscopic model, we study the electric and
magnetic properties of the interface between a half metallic
manganite and an insulator.  We find that the lack of carriers at
the interface debilitates the double exchange mechanism, weakening
the ferromagnetic coupling between the Mn ions. In this situation
the ferromagnetic order of the Mn spins near the interface is
unstable against antiferromagnetic CE correlations, and  a
separation between  ferromagnetic/metallic  and
antiferromagnetic/insulator phases at the interfaces can occur. We
obtain that the insertion of extra  layers of undoped manganite  at
the interface introduces extra carriers  which reinforce the double
exchange mechanism and suppress antiferromagnetic  instabilities.

\end{abstract}


\maketitle

\section{Introduction}
Half metallic ferromagnets are materials in which the electronic
carriers at the Fermi energy have all the same spin direction. They
are very promising materials for spintronics, since they can be used
as spin polarized current injector and/or detector in
magnetoelectronic devices.

For example, in a magnetic tunneling junction device the relative
orientation of the magnetization of the two ferromagnetic electrodes
affects dramatically the electron transport across the tunneling
barrier connecting them\cite{Moodera_2001,applications}. In these
devices the tunneling magnetoresistance, TMR$=(R_{AP}- R_P)/R_{AP}$,
where $R_{AP}$ and $R_P$ are respectively the resistances for
antiparallel and parallel orientations, is directly related to the
spin polarization of the electrodes by Julli\`{e}re
formula\cite{Julliere,Slonczewski} in such a way that as larger is
the spin polarization of the electrodes, larger is the TMR.
Independently of the details of the barrier,the TMR gets its maximum
value when using half metallic electrodes\cite{Brey_2004b}. Large
values of TMR are desirable for optimal use of magnetic tunneling
junctions in technological applications, and therefore the search
for half metallic materials is one of the more active research area
in solid state science\cite{Coey_2004}.

Although several materials have been proposed to exhibit
half-metallic conductivity, only in a small number of them this
property has been experimentally confirmed. Some of the most studied
half metallic materials are perovskites of manganese of formula
(R$_{1-x}$D$ _x$)MnO$_3$, where R denotes rare earth ions (R=La,
Pr,...) and D is a divalent alkaline ion (D=Ca,Sr,...). These
compounds are called generically manganites. In these oxides  $x$
coincides with the concentration of holes moving in the $e_g$
orbital bands of the Mn ions that ideally form a cubic structure.
For perovskites of the form La$_{1/3}$D$_{2/3}$Mn O$_3$, spin
polarized photoemission experiments\cite{Park_1998} and low
temperature magnetoresistance
measurements\cite{Viret_1997,Bowen_2003} indicate an almost complete
spin polarization of the carriers and  the half metallicity
character of these oxides.

The experimental evidence of the large spin polarization of
manganites, particularly La$_{2/3}$Ca$_{1/3}$MnO$_3$ and
La$_{2/3}$Sr$_{1/3}$MnO$_3$, makes these materials very good
candidates for electrodes in tunnel devices. However, all
experimental studies agree\cite{Yu_1996,Izumi_2001} to show that the
TMR of manganites based devices decays rapidly with temperature and
practically vanished much below room temperature, which in some
manganites is below the Curie temperature (360K for
La$_{2/3}$Sr$_{1/3}$MnO$_3$). The decay  of the TMR with temperature has been ascribed to a
reduction of the spin polarization at the electrode-barrier
interface\cite{Izumi_2001,LeClair_2000,Bibes_2001,Bibes_2002}.

Spin polarization experiments have shown that the magnetization of a
free La$_{2/3}$Sr$_{1/3}$MnO$_3$ surface decreases much more rapidly
with temperature than the spin polarization in bulk
materials\cite{Park_1998}. This degradation has been  attributed to
oxygen deficiency at the manganite surface,  which splits the Mn
$e_g$ orbitals weakening the double exchange mechanism
(DE)\cite{Zener,Anderson,DeGennes} for ferromagnetic (FM) order and
favoring an antiferromagnetic (AF) arrangement of the Mn ions  at
the surface\cite{Calderon_1999}.

The manganite/insulator interface is different than the manganite
surface. Recently, Garcia {\it et al.} \cite{Garcia_2004} studied
thermal decay of the spin polarization of
La$_{2/3}$Sr$_{1/3}$MnO$_3$/insulator interface, and show that the
magnetization of the interface can be as robust as that of the bulk.
This result points out the difference between maganite/insulator
interfaces and manganite surfaces and underscores the importance of
the electronic carriers at the interfaces in order to obtain high
temperature large TMR devices. Also Yamada {\it et
al.}\cite{Yamada_2004} have proved that by grading the doping
profile at the interface, robust ferromagnetic order can be realized
around room temperature. In this direction, Density Functional
Theory based calculations\cite{Zenia_2006} have shown that the
magnetic character of the La$_{0.7}$Sr$_{0.3}$MnO$_3$/SrTiO$_3$
junction depends on the interface termination. The
La$_{0.7}$Sr/TiO$_2$ interface is FM and metallic. However in the
MnO$_2$/SrO interface the density of carriers is smaller and the DE
mechanism is not strong enough to overcome the  AF superexchange
interaction and the interface presents antiferromagnetic order.
Also, recently Lin {\it et al. }\cite{Lin_2006} have studied the
properties of  double exchange superlattices, and predict a rich
phase diagram, where the magnetic phases are correlated with the
electronic charge distribution.

In the case of La$_{2/3}$Ca$_{1/3}$MnO$_3$ films on (001) SrTiO$_3$
substrates, NMR experiments have shown that an electronic phase
separation into conducting and non conducting phases occurs close to
the interface\cite{Bibes_2001}. On the contrary, films grown on
(110) substrates show no trends of electronic phase
separarion\cite{Fontcuberta_2006}. Because the (001) films are fully
strained whereas those  grown  on (110) are partially relaxed, the
existence of electronic phase separation has been attribute to the
strain. The strain induce a transition from a orbitally disordered
ferromagnetic state to an orbitally ordered state associated with
antiferromagnetic stacking of manganese oxides
planes\cite{Klein_2002}. Epitaxial strain is also the responsible of
the appearance of electronic phase separation in
La$_{0.6}$C$_{0.4}$MnO$_3$ films grown on (100)
NdGaO$_3$\cite{Sanchez_2006}.

In this work we analyze theoretically the manganite/insulator
interface. We use a microscopic model  able to describe different
exotic ground states that appear in bulk manganites at particular
electron
concentrations\cite{Brey_2004,Brey_2005,Brey_2005b,Salafranca_2006}.

We assume that the manganite has an ideal cubic perovskite structure
, ABO$_3$. The Mn ions are located in the B sites and form a cubic
lattice. The A sites of the perovskite crystal contain ions with
average charge $(1-x)|e|$. The Mn ions have a core spin $S$=3/2,
created by three electrons located in the deep energy $t_{2g}$
levels. In addition there are $1-x$ electrons per unit cell, that
hop between the $e_g$ orbitals of the Mn ions. The carriers are
coupled to the Mn's core spins through a very large Hund's coupling,
in such a way that the motion of the carriers creates  a long range
ferromagnetic order. This mechanism for ferromagnetism is called
double exchange interaction \cite{Zener,DeGennes,Anderson}. The DE
competes with an AF superexchange coupling between the Mn core
spins, and as result of this competition different exotic ground
states emerge\cite{brink,Arovas-1999,dagottobook,Brey_2004}.

We study a slab of manganite sandwiched into  an insulator. We
consider  (001) manganite/insulator interfaces, where the average
ionic charge changes from its bulk value to zero in a unit cubic
lattice, see Fig.1. The long range Coulomb interaction is taken into
account by the Hartree approximation. The main effect of the
insulator is to confine  the carriers to move in the manganite slab.

The main conclusion of our work is that electronic phase separation
between a FM metallic phase and a spin and orbital ordered insulator
phase is likely to occur at the manganite/insulator interface. This
instability is favored by the reduction of carriers at the interface
which weakens the FM coupling between the Mn ions,  making more
relevant the superexchange AF interaction. The ferromagnetic order
at the interface can be recovered by inserting a small number of
undoped AMnO$_3$ layers at the junction. Extra layers of AMnO$_3$ at
the interface  supply carriers to adjacent manganite slabs favoring
a ferromagnetic interface.

The rest of the paper is organized in the following way; In Section
II we introduce the microscopic model and we discuss the more
relevant electronic and magnetic phases that appear in bulk
manganites. In Section III we show the electronic and magnetic
phases that appear at the manganite/insulator interface as function
of the strength of the long range Coulomb interaction and the
superexchange AF coupling between the Mn ions. We study the case of
a simple manganite/insulator interface and also  we analyze how the
properties of the interface  are modified when  a layer of undoped
manganite is introduced at the interface. In Section IV we summarize
our results.

\begin{figure}
  \includegraphics[clip,width=9cm]{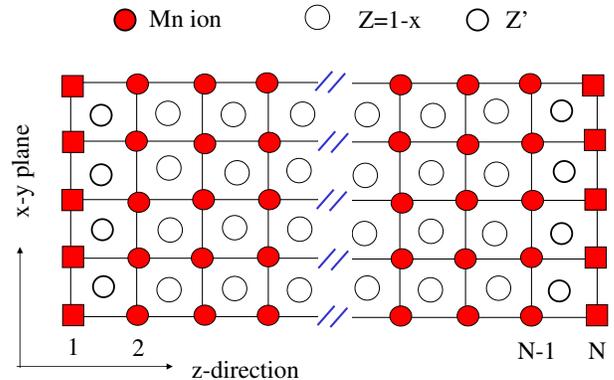}
  \caption{(Color online) Schematic representation of the heterostructure studied in this paper. We choose
  the $x-y$ plane as the interface and $z$ as the growth direction.
  The red circle denote the Mn ions. The green shaded and open circles represent ions of charge
  $Z=1-x$ and $Z'$ respectively. The red squares show the sites where, because of
  the
  confinement by the insulator,  the
  wavefunctions of the carriers has to be zero.
  }
\label{fig1}
\end{figure}

\section{Model}
{\it Heterostructure geometry.} In the cubic perovskites the Mn ions
are located at positions $\textbf{R}_i = a (n_i,m_i,z_i)$ and form a
cubic structure of lattice constant $a$. The positions A of the
perovskite AMnO$_3$, $\textbf{R}^A _i = a (n_i +1/2,m_i +1/2,z'_i
+1/2)$,  are occupied, in the proportion R$_{1-x}$D$_x$, either by a
rare ion ,R, or by a a divalent alkaline ion ,D. The divalent and
trivalent atoms have a ionic charge +1 and 0 respectively, with
relation to the Mn site. In our model we consider a (100) AMnO$_3$
slab containing N planes of MnO$_2$ intercalated by N-1 planes of
AO. We consider periodic boundary conditions in the $x$ and $y$
directions. In the $z$-direction the positions of the Mn ions run
from   $z_i  =1 $ to $z_i  =N,  $ whereas the positions of the
counterions run from $z_i'=0$   to $z_i'=N-1$.

This slab is confined by a wide gap insulator,e.g. TiSrO$_3$. The insulator  is described in our model by
infinity potential barriers that confine the itinerant carriers to move in the manganite slab.

In order to simulate two symmetric manganite/insulator interface, we
locate in the two outmost AO planes atoms with an ionic charge $Z'$.
The $Z'$=0 case corresponds to a divalent atom and represents an
interface with a deficit of electron charge. The $Z'$=1 case
corresponds to a trivalent ion and in this situation there is an
excess of electron carriers at the interface. In the central part of
the slab we describe a manganite of formula R$_{1-x}$D$_x$MnO$_3$ by
locating in the sites A of the crystal structure virtual ions of
charge $Z |e|=(1-x)|e|$. The charge neutrality requires to get an
electron concentration per Mn ion of [(N-3)(1-x)+2$Z'$]/N.

With this geometry the insulator confines the carriers to move in the manganite region, and the long range
Coulomb interaction forces
the electronic charge  to follow the spatial distribution of the charged ions.

{\it Microscopic Hamiltonian.} In the manganites the crystal field
splits the Mn $d$ levels into a fully occupied strongly localized
$t_{2g}$ triplet and a doublet of $e_g$ symmetry. For manganites of
composition R$_{1-x}$D$_x$MnO$_3$, there are $1-x$ electrons per Mn
that hop between the $e_g$ $Mn$ states. The Coulomb interaction
between electrons prevents double occupancy and aligns the spins of
the $d$ orbitals.  The Hund's coupling between the spins of the
carriers and each core spin is much larger than any other energy in
the system, and each electron spin is forced to align locally with
the core spin texture. Then the carriers can be treated as spinless
particles and the  hopping amplitude between two Mn ions is
modulated by the spin reduction factor,
\begin{equation}
f_{12}= \cos\frac{\vartheta _1}{2}\cos\frac{\vartheta _2}{2} + e
^{i ( \phi _1 - \phi _2)} \sin\frac{\vartheta
_1}{2}\sin\frac{\vartheta _2}{2} \label{SRF}
\end{equation} where
$\vartheta _i$ and $ \phi _i $ are the polar and azimuthal angles
Euler angles of the, assumed classical, Mn core spin $\textbf{S} _i
$ . This is the so called  DE model\cite{Zener,Anderson,DeGennes}.

The microscopic  Hamiltonian we study has the following terms,
\begin{equation}
H=H_{DE}+ H_{AF}+H_{U'}+ H_{Coul}\, \, \,
\label{Total_H}
\end{equation}

The first term, $H_{DE}$ describes the motion of the carriers,
\begin{equation}\label{H_DE}
 H_{DE}   =   -\sum_{i,j,a,a'} f _{i,j} \, t _{a,a'} ^{u} \,  C ^+ _{i,a} C
_{j,a'} \, \, \, ,
\end{equation}
where $C^+ _{i,a}$ creates an electron in  the Mn ions located at site $i$ in the $e_g$ orbital $a$ ($a$=1,2 with
1=$|x^2-y^2>$ and 2=$|3z^2-r^2>$). The hopping amplitude is finite for next neighbors  Mn and depends both on
the type of orbital involved and on the direction $u$ between sites $i$ and $j$ ($t_{1,1}^{x(y)}=\pm \sqrt{3}
t_{1,2}^{x(y)} =\pm \sqrt{3} t_{2,1}^{x(y)}=3t_{2,2}^{x(y)}=t)$\cite{dagottobook}. Along this work $t$ is taken as the energy
unit.

The second term describes the AF coupling between first neighbors Mn core spins,
\begin{equation}
\label{H_AF}
H_{AF} =J_{AF} \sum _{<i,j>}
\textbf{S} _i \textbf{S} _j \, \, \, ,
\end{equation}
being $J_{AF}$ the strength of the antiferromagnetic coupling between the Mn core spins.

In Eq.\ref{Total_H}, the term
\begin{equation}
\label{H_U}
H_{U'} = U' \sum _i \sum_{a \neq a'} n _{i,a}n _{i,a'} \, \, \, \, ,
\end{equation}
with $n_{i,a} \equiv C^+ _{i,a} \,C _{i,a}$, is a repulsive interaction between electrons in the same ion,
but when they are in different orbitals.

Finally, the last term of Eq.\ref{Total_H} describe the long-range Coulomb interaction between
the charges in the system.
\begin{eqnarray} \label{H_Coul}
H_{Coul}  =  \frac {e ^2}{\epsilon} \sum _{i \neq j}& &  \left  [
\frac { 1 }{2 } \frac {  <n_i> <n_j>}{|\textbf{R}_i-\textbf{R}_j|}
+\frac {1}{2 } \frac {  Z_i Z_j }{|\textbf{R}_i ^A -\textbf{R}_j
^A|} \right. \nonumber
\\  & & \left .
  -  \frac { Z_j <n_i>}{ |\textbf{R}_i  -\textbf{R}_j ^A|} \right ]\, \, \, ,
\end{eqnarray}
here  $< \! \! n_i \! \! > = \sum _{a}< \! C^+ _{i,a} \,C _{i,a} \! >$ is the occupation number of the
Mn ion locates at site $\textbf{R} _i$, $Z_i e$ is the average charge of the ion locate at sites $\textbf{R}_i ^A$ and
$\epsilon$ is the dielectric constant of the material.

{\it Energy scales and mixing terms.} In this model there are four
energy scales, the hopping amplitude $t$, the AF coupling between Mn
ions, the Hubbard term $U'$ and the screening parameter, $\alpha = e
^2/a\epsilon t$, which measures the strength of Coulomb
interaction\cite{Lin_2006}. The manganites with a FM metallic ground
state are characterized by a bandwidth of order 4-6$eV$ and a
background dielectric constant $\epsilon \sim 5-10$ implying a value
of $\alpha$ in the range 0.7-2. The AF coupling is the smallest
parameter, which according to some estimations, is $J_{AF}\sim
0.1t$\cite{Perring_1997,dagottobook}. Even with this small value
$J_{AF}$ play a fundamental role in stabilizing interesting
experimentally observed phases of
manganites\cite{brink,dagottobook,Brey_2005}. The Hubbard term $U'$
has been included in the Hamiltonian for stabilizing, at moderate
values of $J_{AF}$, the magnetic A phase at $x$=1. In the $A$ phase,
a  Mn spin  is   ferromagnetically  coupled with the Mn spins
located in the same plane ($x-y$), and antiferromagnetically with
the Mn spins belonging to different planes. The value  $U'$=2$t$,
stabilizes  the $A$ phase at $x=1$, and  does not affect the phase
diagram at other hole concentrations.

A relevant contribution to the Hamiltonian which has been not
included in Eq.\ref{Total_H}, comes from the lattice deformation. In
orbital ordered phases, it is possible to reduce considerable the
energy by coupling the orbital order with the Jahn-Teller
deformation of the oxygen octahedra surrounding the Mn ions. In
large extent the lattice contribution of the ground state energies
can be described by using effective values of $J_{AF}$\cite{brink},
and therefore the value of the AF coupling that we use in our model
may be larger than the expected from magnetic neutron scattering
experiments\cite{Perring_1997}.

For a given value of the parameters $U'$, $\alpha$ and  $J_{AF}$,
and a texture of core spins $\{ \textbf{S}_i \}$, we solve
self-consistently the mean field version of Hamiltonian
(\ref{Total_H}) and obtain the energy, the local charges
$\{\rho_i\}$ and  the orbital order $\{\tau _{xi}, \tau_{zi} \}$.
The orbital order is characterized by the expectation value of the
$x$ and $z$ components of the orbital pseudospin, $\tau _{xi}= C^+
_{i1}C_{i2} + C^+ _{i2} C_{i1}$ and $\tau _{zi}= C^+ _{i1}C_{i1} -
C^+ _{i2} C_{i2}$, respectively.

\subsection{Relevant bulk phases.}
The ground state properties of manganites are determined by the
competition between the  energy scales: $J_{AF}$, $U'$, $t$,
$\alpha$. In manganites, the energies involved in these interactions
are comparable  so very different states can have very similar
energies. That is the reason why by slightly varying parameters such
as carrier concentration, strain, disorder, or temperature,
different ground states such as ferromagnetic metallic
phases\cite{Wollan}, AF Mott insulator\cite{Kanamori}, charge and
orbital ordered stripe phases\cite{Chen1,Chen2,Mori,Chen3}, or
ferromagnetic charge ordered phases\cite{Loudon1} can be
experimentally observed. We have checked that the Hamiltonian
Eq.\ref{Total_H}  with the appropriated parameters describes  the
bulk ground states previously
reported\cite{dagottobook,Brey_2004,brink,Solovyev}.

\begin{figure}
  \includegraphics[clip,width=9cm]{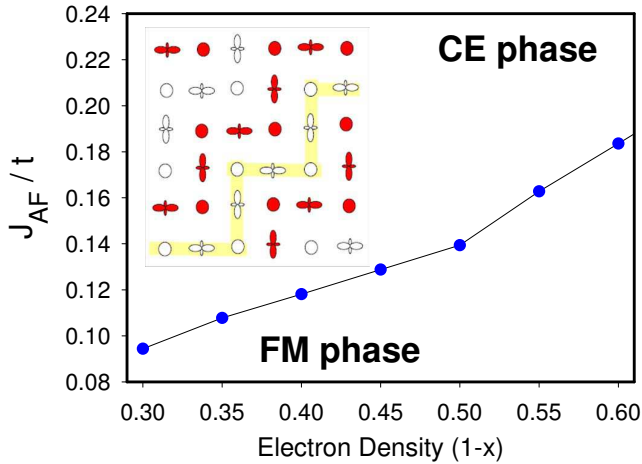}
  \caption{(Color online) Bulk phase diagram $J_{AF} -(1-x)$ for $x$ near $x$=1/2.
  In the inset it is represented the orbital and spin order
  of the CE phase at $x$=1/2. Elongated orbitals along the $x(y)$ directions
  represent  $d_{3z^2-r^2}$($d_{3y^2-r^2}$) orbitals. Circles represent $d_{x^2-y^2}$ orbitals.
  Red and white symbols indicate up and down Mn spins respectively. Shaded regions emphasize the CE zigzag chain.
  The line separating the FM phase from the CE phase is practically independent of $U'$ and
  $\alpha$ for realistic values of these parameters ($U' <4t$, and $\alpha <2t)$.
  }
\label{fig2}
\end{figure}

At the manganite/insulator interface, the electron concentration
falls from $1-x$ to zero in a couple of lattice parameters. In order
to analyze  the  electronic reconstructions at the interface, it is
convenient to study the bulk properties as function of the electron
concentration. In Fig.2 we plot the $J_{AF} -(1-x)$ phase diagram
near $x=1/2$. Here we only consider the main two phases: the FM
metallic phase and the CE phase. In the CE phase, the $x$-$y$ layers
are  AF coupled while the  magnetic structure  within the planes is
that of ferromagnetic zig-zag chains coupled
antiferromagnetically\cite{Wollan}. The horizontal ($x$) and
vertical ($y$) steps of the zig-zag chains contain three Mn ions. In
the CE phase the charge is stacked in the $z$-direction whereas in
the $x$-$y$ planes  it is ordered in a checkerboard form.  Due to
form of the $e_g$ orbitals, the hopping amplitude between two Mn
depends on the type of orbitals involved and on the direction of the
vector linking the ions. In the CE phase, the system takes advantage
of this effect, and creates an orbital ordering along the zig-zag
chain, in such a way that the system opens a gap at the Fermi
energy\cite{Brey_2005}. The orbital ordering is described by a
uniform $z$-orbital pseudospin, $\tau_{z,i}$ and  two finite Fourier
components of the $x$-component of the pseudospin, $\hat{\tau
_x}(\pi/2,\pi/2)=\hat{\tau _x}(-\pi/2,-\pi/2)\neq 0$. In the CE
phase the manganites are band insulators and the gap is created by
the orbital order. Note that  electron charge modulation is only
present for Jahn-Teller coupling different from zero. The cusp at
$x$=1/2 in the line separating the FM phase from the CE phase
indicates the existence of the electronic gap.  Note in
Fig.\ref{fig2} that the critical $J_{AF}$ value for the FM to CE
transition increases with the electron concentration. This is
because the kinetic energy of the metallic FM phase increases with
electron density, and it is necessary larger AF interaction for
overcoming it.

In Fig.\ref{fig2} we have not consider the A phase or other  more
sophisticated phases\cite{dagottobook,Brey_2005} that can appear
when moving from half filling. These phases are very close in energy
to the CE phase, they  also represent orbital and spin ordered
phases and filled only a small part of the phase diagram. Therefore
for simplicity we ignore them and just consider the competition
between the CE and the FM phase.

\begin{figure}
  \includegraphics[clip,width=9cm]{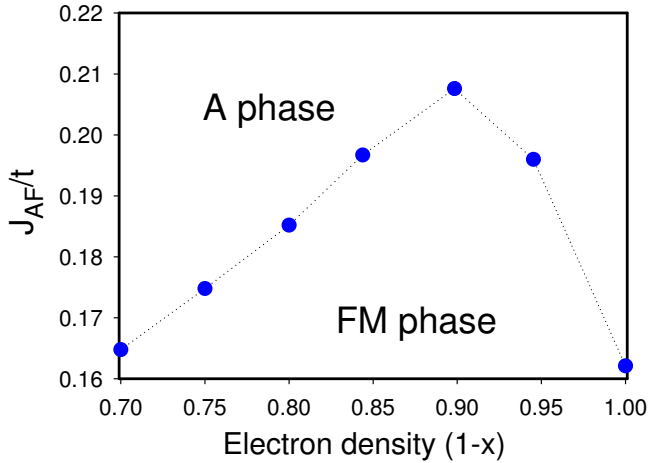}
  \caption{(Color online) Bulk phase diagram $J_{AF} -(1-x)$ for $x$ near $x$=0.
  }
\label{fig3}
\end{figure}

In Section III, we are going to consider the manganite/insulator
interface modified by introducing an undoped manganite layer. In
that case the density of electrons changes from the unity to zero at
the interface. Experimentally\cite{Kimura_2003}, undoped manganites
of small  ionic radius, as LaMnO$_3$, are insulators and have an AF
order type A. In the A phase, a Mn spin is ferromagnetically coupled
with the Mn spins located in the same plane $(x-y)$, and
antiferrimagnetically with the Mn spins belonging to different
planes. In Fig.\ref{fig3} we plot the $J_{AF} -(1-x)$ phase diagram
near $x=0$. By increasing the AF coupling  the system undergoes a
FM-AFM transition at a $x$-dependent critical value $J_{AF} ^C$. For
hole concentration greater than $x$=0.1, $J_{AF} ^C$ increases with
the electron density, reflecting the increase of the kinetic energy.
$J_{AF} ^C$ has a maximum near $x=0.9$, and present a minimum for
$x$=1. This minimum occurs because the A phase is an orbital ordered
phase with a gap at the fermi energy. Near $x$=0, the A phase
competes in energy with the E phase\cite{Salafranca_2006}. Be use  a
Hubbard term, $U'=2t$,  in order to  privilege the A phase, and
describe correctly the ground state of LaMnO$_3$. This value of $U'$
practically unaffect the phase diagram of the manganites near half
doping.

\section{Manganite/Insulator interface. Results}
We have solved  self-consistently the mean field version of
Hamiltonian (\ref{Total_H}) for  manganite/insulator interfaces, see
Fig\ref{fig1}, with different spin textures. We analyze two extreme
cases; the $Z'$=0 case where there is a deficit of charge at the
interface and the $Z'$=1 case where extra carriers are confined at
the interface.

\subsection{$Z'$=0. Defect of electron charge}

\begin{figure}
  \includegraphics[clip,width=9cm]{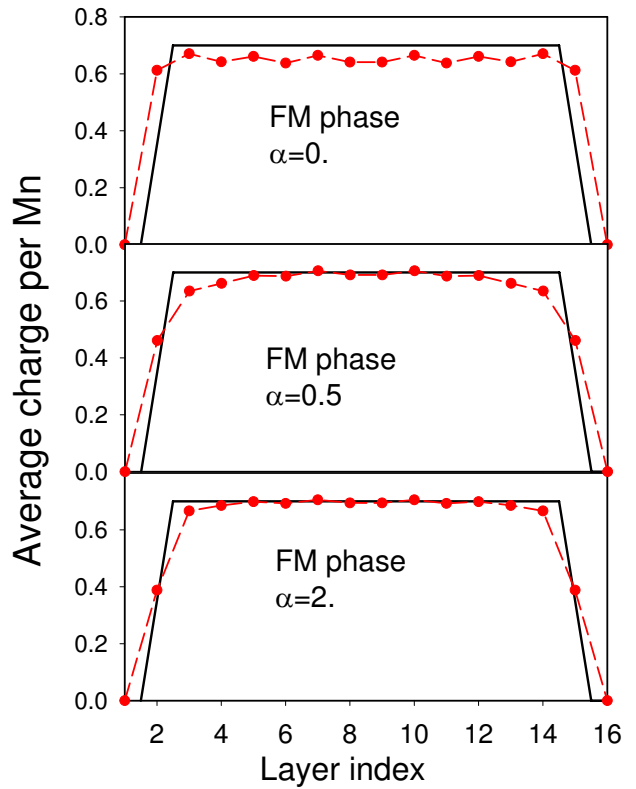}
  \caption{(Color online) Electronic charge on the Mn ions, calculated in the FM phase, for different values of
  the Coulomb interaction $\alpha$. The calculation is done in a
  slab formed by 16 layers of MnO$_2$. The continuous  line
  represents
  the background of positive charge  created by the ions
  located in the AO layers. The electron wavefunctions are forced to
  be zero at the first and last MnO$_2$ layers.
  }
\label{fig4}
\end{figure}

In Fig.\ref{fig4} we plot the electronic charge on the Mn ions in a
FM slab of R$_{0.7}$D$ _{0.3}$MnO$_3$ containing 16 planes of
MnO$_3$. In the case of zero Coulomb interaction ($\alpha$=0) the
electronic charge oscillates slightly around $<n_i> \sim$0.65 that
is the average  charge per electronic active Mn. These fluctuations
are Friedel like oscillations and appear because of the confinement
of the carriers to move into the slab.  In the absence of Coulomb
interaction the charge on Mn ions drops abruptly to zero in just one
lattice spacing.  When increasing the Coulomb interaction the
electronic charge wants to screen the background of positive charge
created by the counterions and approaches the bulk value $<n_i>$=0.7
at the center of the slab. Also, at the interface, the  electronic
charge drops smoother from $<n_i> \sim $ 0.7 to $<n_i>$=0,  as the
Coulomb interaction parameter $\alpha$ increases. For moderate
values of the Coulomb interaction, there is a MnO$_2$ slab at the
manganite/insulator interface where the average electronic charge is
close to $<n_i>$=0.5. Since near half doping the manganites can be
unstable to form CE-type AF order, we also have studied interfaces
where in one or two MnO$_2$ layers, the core spins of the Mn ions
are CE ordered.

\begin{figure}
  \includegraphics[clip,width=9cm]{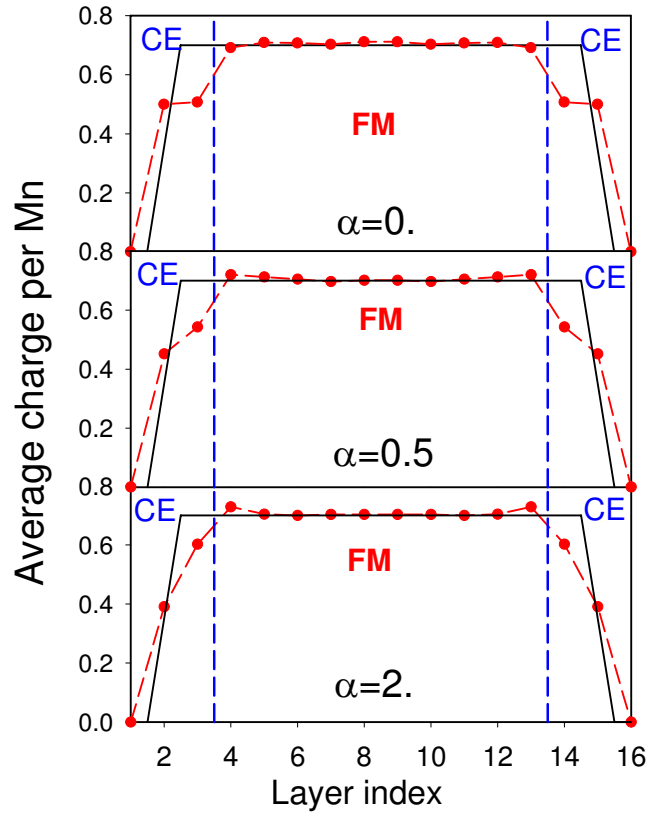}
  \caption{(Color online) Electronic charge on the Mn ions for different values of
  the Coulomb interaction $\alpha$.
  In this calculation  the two first and two last  electronic
  active MnO$_2$ layers  have an antiferromagnetic order of type CE.
  The vertical dashed lines indicate the border between  the FM region  and  the CE layers.
  The calculation is done in a
  slab formed by 16 layers of MnO$_2$. The continuum  line
  represents
  the background of positive charge  created by the ions
  located in the AO layers. The electron wavefunctions are forced to
  be zero at the first and last MnO$_2$ layers.
  }
\label{fig5}
\end{figure}

In Fig.\ref{fig5} we plot the average electronic charge on Mn ion in
a slab of R$_{0.7}$D$ _{0.3}$MnO$_3$ where the two pairs of
electronic active  MnO$_2$ layers closest to the interfaces are
AF-CE ordered. In absence of Coulomb interactions,  the electronic
charge in the CE layers is pinned to a value closer to $<n_i>$=1/2.
These plateaus in the charge density reflects the incompressibility
of the electronic system at half doping in the CE phase, see
Fig.\ref{fig2}. As the Coulomb interaction increases the electron
charge prefers to follow the profile defined by the counterions and
the plateau in the electronic density profile becomes weaker. In any
case, the existence of a region of incompressible electronic density
is evident in the charge density profile even for rather large
Coulomb interaction.

\begin{figure}
  \includegraphics[clip,width=9cm]{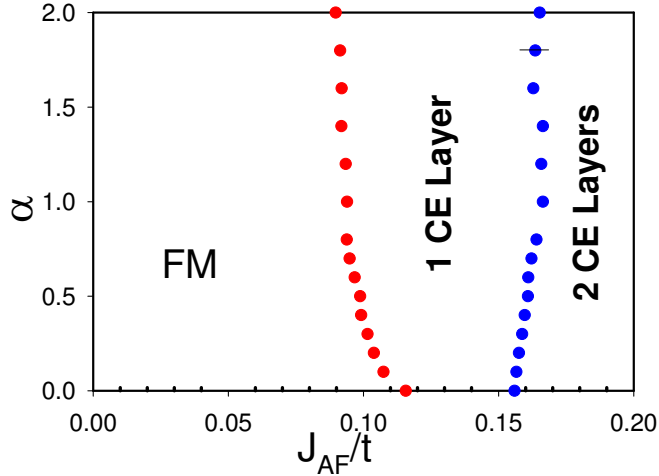}
  \caption{(Color online)  Phase diagram $J_{AF} -\alpha$ for a
  manganite/insulator interface. In bulk, the manganite has a hole
  concentration $x$=0.3. The error bar is an estimation of the
  numerical error in the calculations.
  }
\label{fig6}
\end{figure}

The differences in energy between the FM phase and the phases
containing AF-CE ordered MnO$_2$ layers depend on the Coulomb
interaction, $\alpha$, and the superexchange AF coupling, $J_{AF}$.
In Fig.\ref{fig6} we plot the phase diagram $\alpha$-$J_{AF}$ for a
system  with a bulk hole concentration $x$=0.3. For a fix value of
$\alpha$,   the number of CE layers in the slab near the interface
increases as the AF coupling increases. For larger values of
$J_{AF}$ than those shown in Fig.\ref{fig6}, all the Mn ions in the
slab order in the AF-CE phase.  An important point to note in
Fig.\ref{fig6}, is that the critical value of the AF coupling for
the appearance of an AF-CE MnO$_2$ layer at the manganite/insulator
interface is considerable smaller than the critical value for the
occurrence of the $x$=1/2 AF-CE phase in bulk manganite. In bulk a
value of $J_{AF}$ larger than 0.14$t$ is necessary for the
occurrence of the CE phase at $x$=1/2, whereas at the interface
values of $J_{AF}$ smaller that 0.1$t$ can induce phase separation
between FM metallic phases and AF-CE insulating phase at the
manganite/insulator interface. In Fig.6 we observe that the critical
$J_{AF}$ for the appearance of a CE MnO$_2$ layer decreases slightly
with $\alpha$. This is because as $\alpha$ increases the electron
concentration in the last MnO$_2$ layer decreases, see
Fig.\ref{fig4}, and according with the phase diagram,
Fig.\ref{fig2}, the critical $J_{AF}$ also decreases.

\subsection{$Z'$=1. Excess of electronic charge.}
The phase separation between CE and FM phases at the manganite
/insulator interface that occurs in the $Z'$=0 case is due to the
lack of electronic charge in the last MnO$_2$ planes. In the $Z'$=0
case the electronic charge per Mn  in these planes is close to 0.5,
and  the system is unstable against CE magnetic order. An evident
way to prevent the existence of phase separation is increasing the
electron carrier at the interface. This can be done by inserting a
layer of undoped manganite, LaMnO$_3$, at the
interface\cite{Yamada_2004}. We simulate this layer by locating
trivalent ions, $Z'$=1, in the outmost AO layers and adding the
corresponding extra electrons to the system.

\begin{figure}
  \includegraphics[clip,width=9cm]{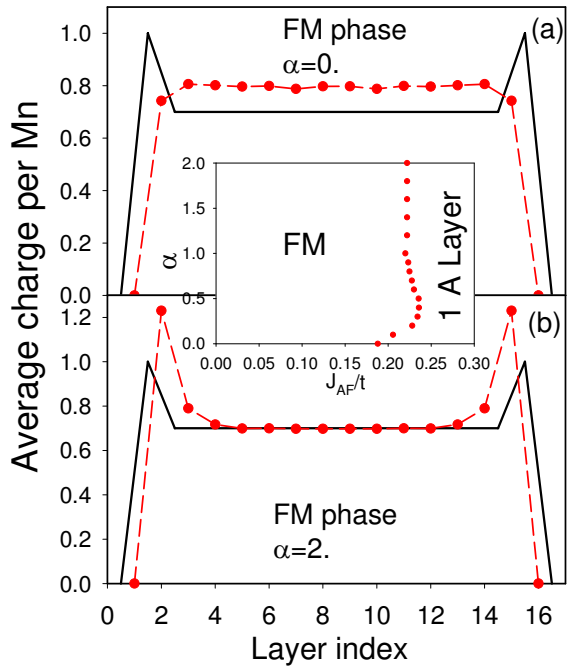}
  \caption{(Color online)Electronic charge on the Mn ions, calculated in the FM phase, (a) for
  $\alpha$=0 and (b) for $\alpha$=2.  The calculation is done in a
  slab formed by 16 layers of MnO$_2$. The continuous  line
  represents
  the background of positive charge  created by the ions
  located in the AO layers, $Z'$=1. The electron wavefunctions are forced to
  be zero at the first and last MnO$_2$ layers. The inset is the
  phase diagram $\alpha$-$J_{AF}$. For realist values of the AF
  coupling, the core spins of all the electronic active  Mn are ferromagnetically ordered.
  }
\label{fig7}
\end{figure}

In Fig.\ref{fig7}, we plot the electronic charge on the Mn ions in a
slab of AMnO$_3$, where the 13 central planes of AO contain
counterions of average charge $(1-x)|e|$ and the two extremal AO
planes contains trivalent ions of charge $Z'$=1. We assume that the
Mn ions are ferromagnetically ordered. Fig.\ref{fig7}(a) corresponds
to the zero Coulomb interaction case, $\alpha$=0. The electronic
charge oscillates near the value of the average charge per
electronic active Mn, $<n_i> \sim$0.79. When the Coulomb interaction
increases, Fig.\ref{fig7}(b), the electronic charge wants to follow
the background of positive charge created  by the counterions. For
large enough values of the Coulomb interaction ($\alpha$=2), the
electronic charge in the central layer of the slab gets its bulk
value $<n_i> \sim$0.7. The excess of charge is located at the
interface, and the charge per Mn ion in the outmost MnO$_2$ layers
is larger than the unity. This excess of charge  prevents the
instability against CE magnetic order and phase separation.  The
insertion of a LaMnO$_3$ layer at the manganite/insulator interface
increases the electron density at the junction , reinforces the DE
mechanism and suppress AF instabilities at the interface.

In Fig.\ref{fig7} we can see that in  the outmost layers the
electron concentration is close  to unity, and the system could be
unstable to flip the spin of the  MnO$_2$ layers close to the
interface and  form A-like AF structures. We have compared the
energy of the FM state with the energy of a phase where the Mn core
spins of the two extreme electronic active MnO$_2$ layers are
antiparallel to the bulk FM polarization. In the inset of
Fig.\ref{fig7} we plot the phase diagram $\alpha - J_{AF}$ for the
slab described above. The main message of this results is that for
realistic values of the AF coupling and Coulomb interaction, the
system is FM.

\section{Summary}

We have studied the magnetic and electronic properties of an
manganite/insulator interface.  We analyze  (001) interfaces, where
the average ionic charge changes from its bulk value to zero in a
unit cubic lattice.  We model the manganites  with  a realistic
microscopic model that describes adequately the different electric
and magnetic phases experimentally observed in bulk. The long range
Coulomb interaction is taken into account by the Hartree
approximation. The main effect of the insulator is to confine  the
carriers to move in the manganite slab. We find that a electronic
phase separation between a FM metallic phase and a spin and orbital
ordered insulator phase is likely to occur at the
manganite/insulator interface. This instability is favored by the
reduction of carriers at the interface which weakens the FM coupling
between the Mn ions,  making more relevant the superexchange AF
interaction. The insertion of a LaMnO$_3$ layer at the
manganite/insulator interface introduces extra carriers at the
junction which reinforce the DE mechanism and the FM long range
order,  suppressing  AF instabilities at the interfaces.

\section{Acknowledgments}
The author thanks J.Salafranca, J.Santamaria  and I.C.Infante for
helpful discussions.  Financial support is acknowledged from Grant
No MAT2006-03741. (Spain).

%\bibliography{mia}

\end{document}